\documentclass[11pt]{article}

\pdfoutput=1

\usepackage[pdftex]{color}
\usepackage{amssymb}
\usepackage{amsthm}
\usepackage{amsmath}
\usepackage{latexsym}
\usepackage{amscd}
\usepackage{graphicx}
\usepackage[pdftex, colorlinks=true, citecolor=green]{hyperref}
\usepackage{lscape}
\usepackage{setspace}
\usepackage{multirow}
\usepackage[table]{xcolor}

\newcommand{\ds}{\displaystyle}
\newcommand{\ben}{\begin{equation}}     
\newcommand{\eeqn}{\end{equation}}
\newcommand{\bey}{\begin{eqnarray}}
\newcommand{\eey}{\end{eqnarray}}
\usepackage{footmisc}

\begin{document}
\begin{flushleft}
\vspace{4mm}
{\bf {\Large A model of predation and survival\\
 in a system of three interacting species}}
\\
\vspace{4mm}
\noindent {\large Anca R\v{a}dulescu$^{*,1}$,
Richard Halpern$^2$, Drew Kozlowski$^1$, Conor O'Riordan$^3$}
\vspace{4 mm}

\noindent $^1$ Department of Mathematics, State University of New York at New Paltz; New York, USA; Phone: (845) 257-3532; Email: radulesa@newpaltz.edu; 

\vspace{2mm}
\noindent $^2$ Department of Physics, State University of New York at New Paltz; New York, USA;

\vspace{2mm}
$^3$ Department of Computer Science, State University of New York at New Paltz; New York, USA

\noindent *Corresponding author

\end{flushleft}

\begin{abstract}
   The study of interactions between multiple species in an ecosystem is an active and impactful direction of inquiry. This is true in particular for fragile systems in which even small perturbations of their functional parameters can produce dramatic effects like species endangerment or extinction, leading the system to enter an unsustainable regime and eventually collapse. In this context, it is important to understand which factors can lead to such effects and for which systems, so that one can act proactively and timely to prevent them. We built and studied a mathematical model that captures the natural interactions between three species, in which two species are predators of the third, but such that one of the predators also consumes the other (to which we refer as Owls, Snakes and Mice). The nonlinear components of the model were documented on existing literature and assembled as a system of Lotka-Volterra ordinary differential equations. Our analytical computations and numerical exploration explorations revealed sequences of transcritical and Hopf bifurcations that underlie counterintuitive transitions of the system into regions of vulnerability to external noise. We conclude that, in order to avoid extinction,one needs to rigorously prescribe a well-documented, prediction-based approach to population control.
\end{abstract} 

\section{Introduction}

 Scientific research in all fields has been pointing out emphatically over the past few decades the many ways in which our environment is vulnerable to change. If one views environmental health as robust coexistence of a diversity of species, vulnerability of such a system may manifest as loss of this robustness, and a tendency to extinction and diminishing diversity. In order to sustainably maintain a long-term healthy regime, eco-systems may have to evolve complex and fine-tuned behaviors, emerging from the complicated interactions between their many components. Then even small external changes imposed on these very well-oiled machines may produce dramatic effects, disturbing their dynamics beyond the point where the known behavior can be restored. Mathematical research over the past two decades has been able to provide efficient modeling frameworks that capture the optimal balance between an eco-system's robustness and vulnerability to both internal and external factors.       

One prevalent topic among the discussion of environmental health is species population. Population concerns are typically associated with a rising number of endangered species. While there are clear direct causes that may contribute to this effect (such as diminishing natural resources, or constant hunting), there are also indirect, less intuitive contributing factors, which are often overlooked and thus remain hard to quantify and address. One of this factors, for example, is overpopulation. The 2013 study by Ghosh and Kar \cite{ghosh2013possible} explores scenarios such as this. Overpopulation threatens not just of one species but all others who directly rely on it. All species are subject to a maximum population limit also known as a caring capacity. When a species exceeds their caring capacity the negative impacts on the species out weigh the positive, leading to extinction. When this occurs, the extinction of one species can often lead to the extinction of another who were positively impacted by the now extinct species. Whether it is for food or for regulating inter-species competition, the absence of the now extinct species can lead to immense population growth or to starvation. Being able to accurately model such interactions  mathematically, offers an important insight to at risk species in an environment. In our paper, we continue in this spirit, and we use a simple model to bring forward a few other counter-intuitive mechanisms that may be responsible for collapse of the eco-system, even in the context of ``sensible'' human control. 

Mathematical modeling is particularly important and powerful in this case, because ``ecological interactions among populations are very complex and can lead to many paradoxical results''~\cite{pal2019hydra}. For example, a recent model of a five-species eco-system discusses counter-intuitive scenarios where increasing the mortality rate of one species inflates it's caring capacity. In cases such as these, an increased mortality rate accelerates population growth and benefits stability. Often known as the Hydra Effect, the odds of the individual in a species worsen while  potential for the species as a whole grows ~\cite{pal2019hydra}. This is an ecological phenomenon that can approached mathematically, since the external and internal interactions of a species are proportional to population. The ability to mathematically predict the effects of changing mortality rates could be extremely advantageous to conservation efforts, offering a unique insight on how to balance environmental elements to greater stabilize an at risk species.

 When looking at species interaction, it is important to note that inter-species interactions are often mutually beneficial ~\cite{singh2022multi}. A 2022 study of ungulates in the Himalayas revealed that the three species of Ungulate studied mutually benefit each others' habitats and food sources. Though this study only includes herbivores, it allows for speculation that the interactions between predators and prey may also have mutually beneficial aspects. This is significant when comparing the effects one species has on another. While it may seem intuitive that predators profit and prey declines, it is important to acknowledge that in some cases a prey species may benefit from being predated. Whether this is because it reduces species competition, or expands the use of habitat, a mathematical model offers the adeptness to foresee counter-intuitive interactions between species.

Predator-prey relationships are therefore a crucial building block to any eco-system, with their coupled dynamics contributing crucially to the system's health and long-term outcome. The traditional framework used for predator-prey interactions has been the Lotka-Volterra model, originally built for two, but later expanded for many  interacting species. While other work had transcended the Lotka-Volterra system to mathematically incorporate more specific inter-species dynamics for their respective case studies, in this paper we will use the original model for three interacting species. We do so, because we want to maintain generality, while we want to keep the model simple~\cite{chauvet2002lotka}.

Three-species Lotka-Volterra systems have been amply studied before, for a variety of schemes, including feed-forward systems with an intermediate predator
~\cite{chauvet2002lotka,pal2024fear} and systems with one predator two pray or one pray two predators~\cite{korobeinikov1999global}. In this paper, we chose to focus on a three-species network formed of a herbivore (represented by mice $M$ in our working example), an intermediate predator which feeds solely off the herbivore (snakes $S$ in our case) and an apex predator which predates on both other two species (owls $O$ in our example), While, in order to fix our ideas, in this paper we will refer to the three species as $M$, $S$ and $O$ (from our working example), this system is representative of any species triplet that satisfies the same predation relationships.

For this layered system, we can then study the impact of various factors, in particular: reproduction and predation rates, access to environmental resources (which may affect intra-specific competition). Most importantly, we also want to investigate and document the impact of human actions and environmental control measures (such as hunting, or deliberate extermination of species which are perceived as detrimental to humans). These various factors will be documented using bifurcation graphs to properly display each parameters effect on each individual species. Each factors will then be analyzed and explained in further detail, in what it means for the system.     

\subsection{Our model}

\begin{figure}[h!]
\begin{center}
\includegraphics[width=0.3\textwidth]{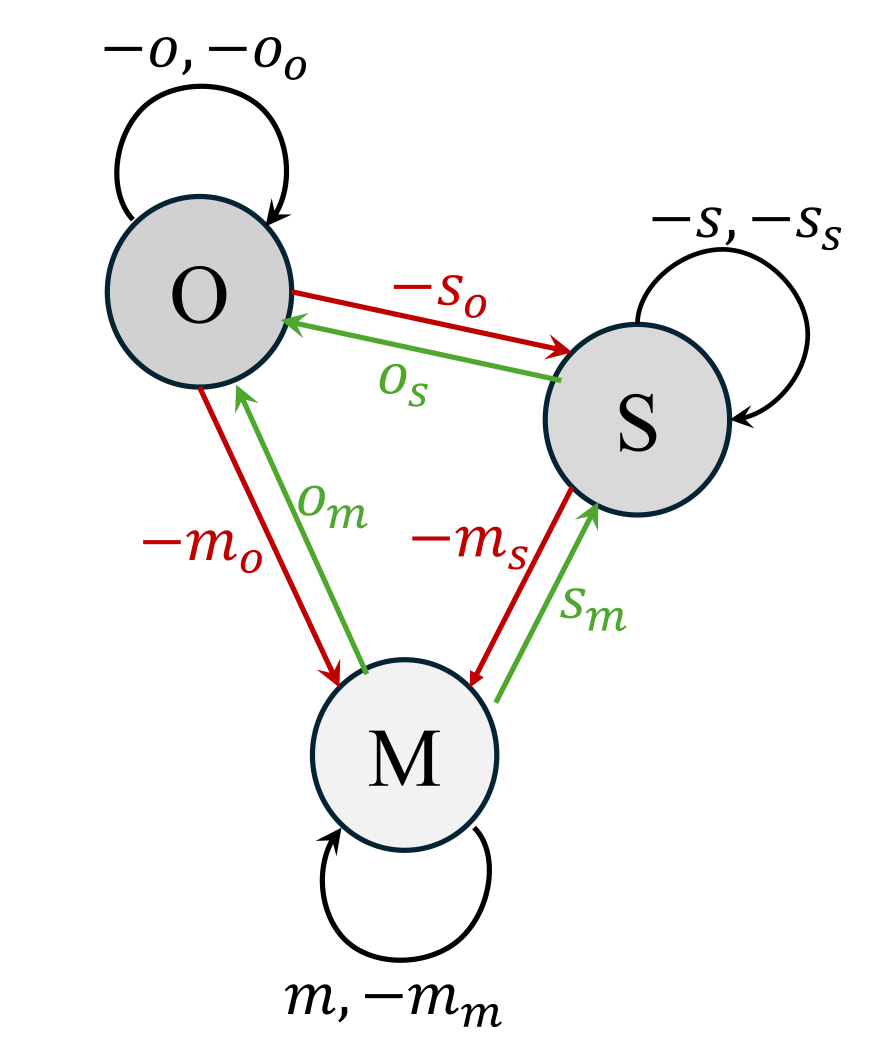}
\caption{\small \emph{{\bf Shematics interactions} between the three species in the system. Red arrows into each node represent predation, and green arrows into each node represent food. The corresponding coefficients are markes on each arrow, with the respective sign. The self-loops represent the birth/death and competitive terms.}}
\label{schema}
\end{center}
\end{figure}

\noindent We study survival and extinction in a system of three predator-prey populations consisting of one prey and two predator species (of which one is the apex predator and will prey on the other, as shown in Figure~\ref{schema}. To fix our ideas and notation, we will refer to these species as owls, snakes and mice, described by the coupled time variables $O$ (owls), $S$ (snakes) and $M$ (mice, respectively). These names are used, however, only for reference, and the system analysis can be generalized to any other system with ternary interactions that have similar predator-prey coupling (such as, for example, lions, hyenas and antelopes). To maintain generality and simplicity, our model considered quadratic terms for all interactions, as follows:

\begin{eqnarray*}
\dot{O} &=& O(o_s S + o_m M - o_o O - o)\\
\dot{S} &=& S(s_m M - s_o O - s_s S - s)\\
\dot{M} &=& M(m - m_m M - m_o O - m_s S)
\end{eqnarray*}

\noindent In absence of prey (i.e., snakes and mice), the owls die out, subject to natural death (at a net rate $-oO$ proportional to the population size) and to interspecies competition (described by the term $-o_o O^2$), as per the traditional logistic model. The owls' survival depends on their successful feeding on snakes ($o_s SO$) and on mice ($o_mMO$), and is not subjected to any predation from other species. In turn, snakes feed on mice ($s_mMS$), and are consumed by owls ($-s_oOS$), and would die out in the absence of both their prey and predators, due to natural causes ($-sS$), as well as internal competition ($-s_sS^2$). Finally, mice are self-sustaining (positive reproduction versus death rate $mM$), but are diminished by both owl and snake predation ($-m_oOM$ and $-m_sSM$, respectively), as well as by internal competition ($-m_mM$). In the absence of any predators, the murine population will stabilize to the carrying capacity $m/m_m$.

The values of the fixed parameters and the ranges of the key parameters are listed in Table~\ref{params}, together with their significance and units. While a more extensive analysis is of course possible, we decided to focus in this paper on dicussing the implications of controlling the reproduction and death rates of the three species (specifically $o$, $s$ and $m$) on the health and viability of the eco-system as a whole. This sprouts from our interest to understand the implications of human control on such a system, control which most often translates into either promoting reproduction of a species or enhancing extermination. 

\begin{table}[h!]
\begin{center}
\begin{tabular}{|l|l|l|l|}
\hline
Variable/parameter & Significance & Value/range & Units\\
\hline
$O$, $S$, $M$ & Number of owls/snakes/mice & 0-100 & $10^3$ individuals (TI)\\
\hline
$o$, $s$, $m$ & Owl/Snake/Mouse reproduction rate & 0-10 & day$^{-1}$\\
$o_o$ & Owl competition rate & 0.5 & TI$^{-1}$day$^{-1}$\\
$o_s$ & Snakes food-value for owls & 1 & TI$^{-1}$day$^{-1}$\\
$o_m$ & Mouse food-value for owls & 0.1 & TI$^{-1}$day$^{-1}$\\
$s_s$ & Snake competition rate & 0.5 & TI$^{-1}$day$^{-1}$\\
$s_o$ & Owl predation efficiency of snakes & 1 & TI$^{-1}$day$^{-1}$\\
$s_m$ & Mouse food-value for snakes & 0.3 & TI$^{-1}$day$^{-1}$\\
$m_m$ & Mouse competition rate & 0.05 \& 0.5 & TI$^{-1}$day$^{-1}$\\
$m_o$ & Owl predation efficiency on mice & 4 & TI$^{-1}$day$^{-1}$\\
$m_s$ & Snake predation efficiency on mice & 1 & TI$^{-1}$day$^{-1}$\\
\hline
\end{tabular}
\end{center}
\noindent \caption{\emph{{\bf System variables and parameters}, with their significance, values and units. In the case of the key parameters, we are showing the ranges we considered in our numerical investigation.}}
\label{params}
\end{table}

In our working example in particular, mice are notoriously exterminated as a pest, snakes may be eliminated in ihnabited areas simply out of fear (likely to a lesser extent than mice), and owls (and other similar predators) are often hunted for sport (or to minimize attacks on poultry in rural areas). Assumptions are often made that a moderate and planned extermination would not impact the system as dramatically as more massive and arbitraty action. For example, one may intuitively feel based on field observation that mice have been reproducing extremely fast, and may consider a moderate level of control appropriate and safe for the integrity of all the species in the eco-system. This consideration may even be based on a comparison with implementation of stronger control in the past,  which did not lead to extinction of any of the three species. Our goal in this analysis is to establish if this intuitive assumption is correct in general. To do so, we will track the behavior of the system under perturbations of the key parameters $o$, $s$ and $m$, considering two separate circumstances that differ in the level of murine intra-specific competition (to show that the context, as captured by the other fixed parameters, may have a significant impact on the outcome). We will establish positivity and stability of equilibria, track down bifurcation points with respect to the key parameters. Based on the traditional analysis of the two-dimensional predator-prey (Lotka-Volterra) system, we expect to find Hopf bifurcations with onset and cessation of stable oscillations in the system. We will carry out direct computations for the components that are tractable analytically, then we will complete the illutrations with numerical simulations.

\section{Analytical results}

A traditional analysis of the system reveals that it has six possible positive equilibria, each relevant as an asymptotic attractor within different parameter regions. To estabilsh local stability, we computed the Jacobian matrix in each case from the general form:
\begin{equation*}
{\cal J}(O,S,M) = \left(  \begin{array}{ccc} o_s S + o_m M -2o_o O & o_s O & o_m O\\ 
-s_o S & s_m M - s_o O -2s_s S -s & s_m S\\ 
-m_o M & -m_s M & m-(2m_mM+m_oO+m_sS) \end{array} \right)
\end{equation*}

\noindent {\bf EQ0: $O^*=S^*=M^*=0$ (total exctinction equilibrium). } The eigenvalues of ${\cal J}(0,0,0)$ are $\lambda_1=-o<0$, $\lambda_2=-s<0$, $\lambda_3=m>0$, hence this equilibrium will be a saddle (unstable), irrespective of the parameter values.\\

\noindent {\bf EQ1: $O^*=S^*=0$ (extinction of owls and snakes), $\ds M^*=\frac{m}{m_m}$ (carrying capacity for mice).} The eigenvalues of $\ds {\cal J}\left( 0,0,\frac{m}{m_m} \right)$ are $\ds \lambda_1=-m, \; \lambda_2= \frac{m o_m-o m_m}{m_m}, \; \lambda_3= \frac{m s_m - s m_m}{m_m}$. This equilibirum is stable iff the the system's parameters simulanesouly satisfy: 

$$(i) \; \frac{o}{o_m}>\frac{m}{m_m} \text{ and } (ii) \; \frac{s}{s_m}>\frac{m}{m_m}$$
\noindent For convenience, we will use the prime notation to denote the nagation of a condition; for example, the negatives of conditions $(i)$ and $(ii)$ will be respectively:
$$(i') \; \frac{o}{o_m}<\frac{m}{m_m} \text{ and } (ii') \; \frac{s}{s_m}<\frac{m}{m_m}$$

\noindent {\bf EQ2a: $O^*=0$ (extinction of owls only)}, with the other two coordinates:
$$\ds S^*= \frac{m s_m-s m_m}{m_s s_m + m_m s_s} \text{ and } M^*=\frac{s m_s + m s_s}{m_s s_m + m_m s_s}.$$ 

\noindent While $M^*$ is guaranteed to be positive, positivity of the component $S^*$ is guaranteed when condition $(ii')$ is met. Moreover, the eigenvalues of the Jacobian in this case are $\lambda_1 = -o+o_m M^*+o_s S^*$ and $\lambda_{2,3}$ given by the roots of the quadratic equation 
$$\lambda^2 + (s_s S^* +m_m M^*) \lambda +S^* M^*(s_m m_s + s_s m_m)=0$$

\noindent Notice first that $\lambda_2 +\lambda_3 = -(s_s S^*+m_m M^*)$ and $\lambda_2 \lambda_3 = S^* M^*(s_m m_s +s_s m_m)$, such that $\lambda_2 + \lambda_3<0$ and $\lambda_2 \lambda_3>0$. Hence the real parts of these two eigenvalues are negative (whether they are real or complex conjugate), and the two corresponding directions are attracting for the equilibrium EQ2a. Moreover, also notice that, if condition $(i)$ is met, then:
\begin{eqnarray*}
\lambda_1 &=& -o+o_mM^*+o_sS^* < -\frac{mo_m}{m_m}+o_mM^*+o_sS^*\\\\
&=& -\frac{mo_m}{m_m}+o_m \frac{s m_s + m s_s}{m_s s_m + m_m s_s}+o_sS^*\\\\
&=& \frac{ms_m-sm_m}{m_ms_s+m_ss_m} \cdot \frac{o_sm_m-o_mm_s}{m_m}  = \frac{O^* (o_sm_m-o_mm_s)}{m_m} 
\end{eqnarray*}
Consider in addition the following condition on parameters:
\begin{equation*}
(iii) \quad \frac{o_s}{m_s} < \frac{o_m}{m_m}
\end{equation*}
\noindent Then, if conditions $(i)$, $(ii')$ and $(iii)$ are met, then $\lambda_1<0$, and the equilibrium EQ2a is stable.\\

\noindent {\bf EQ2b: $S^*=0$ (extinction of snakes only)}, with the other two coordinates:
$$\ds O^*= \frac{m o_m-o m_m}{m_o o_m + m_m o_o} \text{ and } M^*=\frac{m o_o + o m_o}{m_o o_m + m_m o_o}$$

\noindent While $M^*$ is guaranteed to be positive, positivity of the component $S^*$ is guaranteed when condition $(i')$ is met. Here, too, $\lambda_1 = -o+o_m M^*+o_s S^*$ and, for the other two eigenvalues, we have $\lambda_2 +\lambda_3 = -(o_o O^*+m_m M^*)$ and $\lambda_2 \lambda_3 = O^* M^*(o_m m_o +o_o m_m)$, such that $\lambda_2 + \lambda_3<0$ and $\lambda_2 \lambda_3>0$. Hence the two corresponding directions are attracting for the equilibrium EQ2b. Moreover, also notice that, if condition $(ii)$ is met, then:
\begin{eqnarray*}
\lambda_1 &=& -s+s_mM^*-s_oO^* < -\frac{ms_m}{m_m}+s_mM^*-s_oO^*\\\\
&=& \frac{mo_m-om_m}{m_mo_o+m_oo_m} \cdot \frac{s_mm_o-s_om_m}{m_m} = \frac{S*(s_mm_o-s_om_m)}{m_m}
\end{eqnarray*}
Consider in addition the following condition:
\begin{equation*}
(iv) \quad \frac{m_o}{m_m} < \frac{s_o}{s_m}
\end{equation*}
\noindent Then, if conditions $(i')$, $(ii)$ and $(iv)$ are met, then $\lambda_1<0$, and the equilibrium EQ2b is stable.\\


\vspace{2mm}
\noindent {\bf EQ3: the non-extinction equilibrium.} This is the equilibrium in which all three species are preserved. It is obtained as a solution $(O^*,S^*,M^*)$ for the linear system:
\begin{equation}
\left\{  \begin{array}{l} 
o_s S + o_m M - o_o O = o\\  
s_m M - s_o O -s_s S = s\\
m_m M + m_o O + m_s S = m
\end{array} \right.
\end{equation}

\noindent and is biologiclaly relevant when it is positive, hence we are interested in obtaining parameter conditions for which $(O^*,S^*,M^*)$ is in the positive octant. One can compute for example:
\begin{equation*}
M^* = \frac{M^*_{top}}{M^*_{bottom}}
\end{equation*}
\noindent where
\begin{eqnarray*}
&&M^*_{top} = (m_ss_o-m_os_s)(om_o+mo_o)-(o_sm_o+m_so_o)(sm_o+ms_o)\\
&&M^*_{bottom} = (m_ss_o-m_os_s)(o_mm_o+m_mo_o)-(o_sm_o+m_so_o)(m_ms_o+s_mm_o)
\end{eqnarray*} 

\noindent Consider the following condition:
$$(v) \quad \frac{m_s}{m_o} < \frac{s_s}{s_o}$$

\noindent If one considers in this condition in addition to $(i)$ and $(ii)$, we have:
\begin{eqnarray*}
M^*_{top} &<& (m_ss_o-m_os_s)\left(\frac{mo_m}{m_m}m_o+mo_o  \right) - (o_sm_o+m_so_o)\left(\frac{ms_m}{m_m}m_o+ms_o  \right) \\
&=& \frac{m}{m_m} [(m_ss_o-m_os_s)(o_mm_o+o_om_m)-(o_sm_o+m_so_o)(s_mm_o+m_ms_o)] \\
&=&  \frac{m}{m_m} [m_os_o(m_so_m-o_sm_m)-m_os_so_mm_o-m_os_so_om_m-o_sm_os_mm_o-m_so_os_mm_o]
\end{eqnarray*}

\noindent If condition $(iii')$ is satisfied, then $M^*_{top}<0$. Similarly, if $(iii')$ and $(iv')$ are met, we also have
\begin{eqnarray*}
M^*_{bottom} &<& \frac{m}{m_m} [(m_ss_o-m_os_s)(o_mm_o+o_om_m)-(o_sm_o+m_so_o)(s_mm_o+m_ms_o)] \\
&=& \left( m_s\frac{s_mm_o}{m_m}-m_os_s \right)(o_mm_o+o_om_m)-\left( \frac{o_mm_s}{m_m}m_o+m_so_o \right)(s_mm_o+m_ms_o) \\
&=& \frac{m_o}{m_m}(m_ss_m-s_sm_o)(o_mm_o+m_mo_o)-\frac{m_s}{m_m}(o_mm_o+m_mo_o)(m_ms_o+s_mm_o)\\
&=&-\frac{m_s}{m_m}(o_mm_o+m_mo_o)(m_os_s+m_ms_o)<0
\end{eqnarray*}

\noindent Hence, if conditions $(i)$, $(ii)$, $(iii')$, $(iv')$ and $(v)$ are satisfied, we have that $M^*>0$. Similar conditions can be obtained to warrantee positivity of $S^*$ and $O^*$. To establish stability at the non-extinction equilibrium, we compute the Jacobian in this case:
\begin{equation}
{\cal J}(O^*,S^*,M^*) = \left(  \begin{array}{ccc} -o_oO^* & o_s O^* & o_m O^*\\ 
-s_o S^* & -s_s S^* & s_m S^*\\ 
-m_o M^* & -m_s M^* & -m_mM^* \end{array} \right)
\end{equation}

\noindent The characteristic polynomial is then:
\begin{eqnarray*}
P(\lambda) &=& \left \lvert   \begin{array}{ccc} \lambda+o_oO^* & -o_s O^* & -o_m O^*\\ 
s_o S^* & \lambda+s_s S^* & -s_m S^*\\ 
m_o M^* & m_s M^* & \lambda+m_mM^* \end{array}  \right \rvert =  
O^*S^*M^* \left \lvert   \begin{array}{ccc} \ds \frac{\lambda}{O^*}+o_o & -o_s & -o_m \\ 
s_o &\ds \frac{\lambda}{S^*}+s_s & -s_m \\ 
m_o & m_s &\ds \frac{\lambda}{M^*}+m_m \end{array}  \right \rvert\\\\
&=& (\lambda+o_oO^*)(\lambda+s_sS^*)(\lambda+m_mM^*)+(o_ss_mm_o - o_mm_ss_o)O^*M^*S^*\\\\
&& \quad \quad +o_mm_oM^*O^* (\lambda+s_sS^*) + m_ss_m(\lambda+o_oO^*) + s_oo_sO^*S^*(\lambda+m_mM^*)\\\\
&=& \lambda^3 + \lambda^2 [o_oO^*+s_sS^*+m_mM^*]\\\\ 
&& \quad \quad +\lambda[(o_os_s+o_ss_o)O^*S^*+(o_om_m+o_mm_o)O^*M^*+(s_sm_s+s_mm_s)S^*M^*]\\\\
&& \quad \quad +O^*M^*S^*[o_os_sm_m-o_mm_ss_o+o_ss_ms_o+o_mm_os_s+m_ss_mo_o+s_oo_sm_m]
\end{eqnarray*}

\noindent Stability calculations involving the Jacobian are more complicated, and some ideas are sketched in the Appendix. Altogether, however, the analysis carried out thus far supports the switch between stable equilibria across different parameter regions, and suggests the presence of transcritical bifurcations are the switching points. 

To fix our ideas at this point, we can focus on specific questions of how the system's behavior evolves when the reproduction rates fluctuate, or when the internal competition is varied, while the other parameters are fixed. Notice that conditions $(i)$ and $(ii)$ can be controlled by the balance in the reproduction rates (for example, a small enough $m$ ensures both). Conditions $(iii)$ and $(iv)$ can be obtained by tuning the murine intra-specific competition $m_m$ (small $m_m$ leads to $(iii)$ and $(iv')$, while large $m_m$ leads to $(iii')$ and $(iv)$). In turn, condition $(v)$ depends on the level of snake competition $s_s$. 

Imagine now that the rate $m$ is small enough so that conditions $(i)$ and $(ii)$ are satisfied, hence EQ1 is positive and stable. When increasing $m$, the two joint conditions will be broken (in an order that depends on the rest of values in the parameter set). This creates two consecutive transcritical bifurcations, in which the EQ2a and EQ2b curves become successively positive, with a simultaneous swap to local stability. 

Let's note that similar analyses of our set of conditions may suggest other sequences of transcritical transitions. Let's also consider the example of decreasing the internal mice competition $m_m$ (which may correspond to more natural resources becoming available to the population, diminishing competition), in a system where owl predation of mice is very high ($m_o$=4 in our table). For large enough $m_m$, conditions $(i)$, $(ii)$, $(iii')$ are satisfied, and $(iv')$ and $(v)$ are also valid, provided $m_o$ is large enough, hence the non-extinction equilibirum is positive and locally stable. When $m_m$ is decreased, one the system will progressively cross regions where extinction equilibria take over. To better understand and illustrate the details in a few such transitions, we will carry out in the next section a battery of simulations, focusing in particular on the contribution of the reproduction/death parameters.

\section{Numerical results}

\subsection{Dependence on murine reproduction rate $m$}

One of the first natural key questions to investigate is how the survival and health of this three-species eco-system depends on the abundance of mice, the common food source for the predator species. In our model, variations in either birth of death rates of mice are captured by changes in their reproductive rate $m>0$ (assuming births outnumber deaths). In these terms, a smaller $m$ may be the result of slower reproduction, but may also reflect higher death rates. Since rodent extermination is a wide spread practice, we choose to focus in our discussion on an interpretation based around human control of rodents in the environment, and understand its potentially unexpected indirect effects on other species that survive on them as a food source.

\begin{figure}[h!]
\begin{center}
\includegraphics[width=0.8\textwidth]{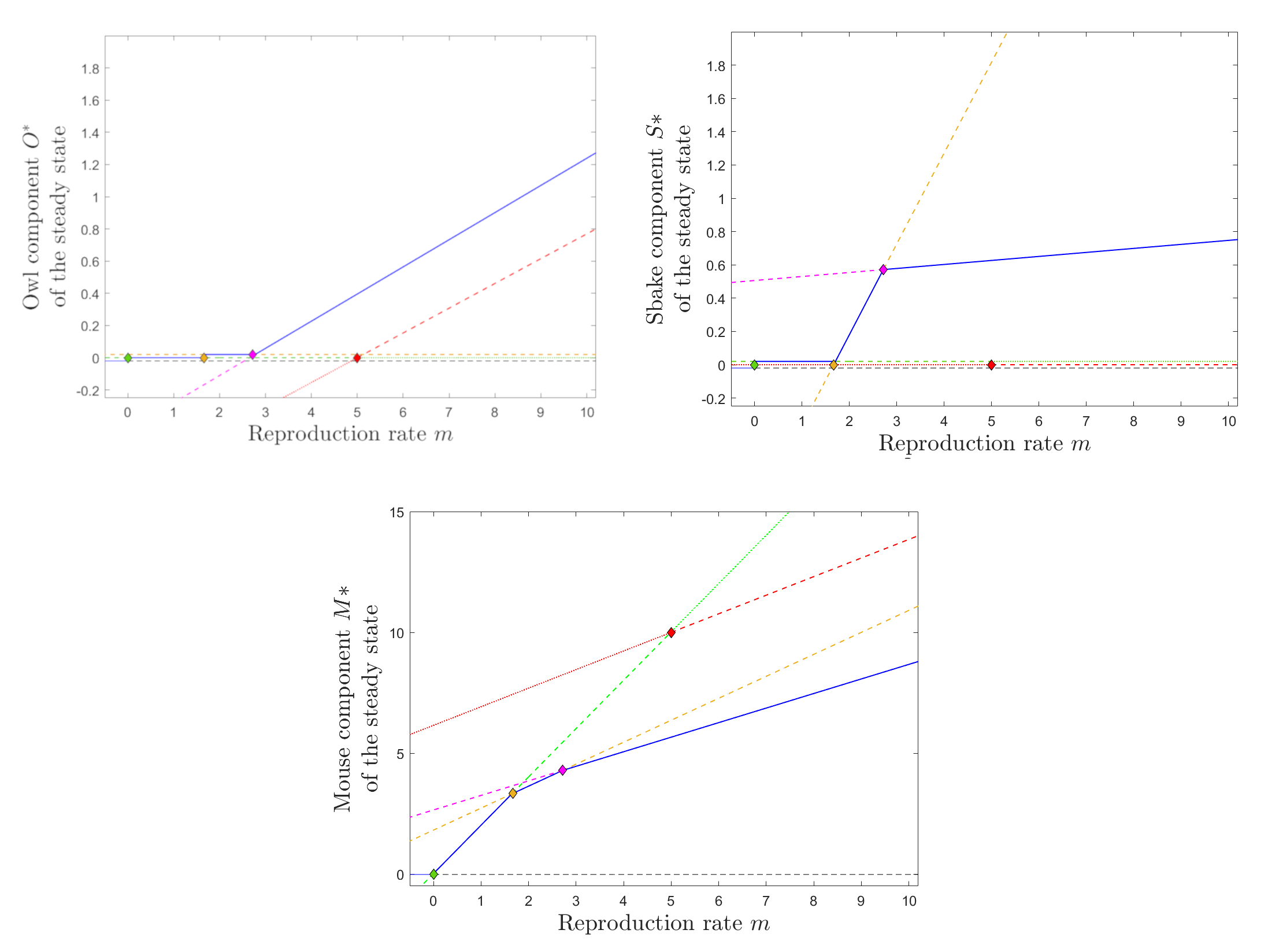}
\caption{\small \emph{{\bf Bifurcation diagram with respect to the mouse reproduction rate $m$}, for high intra-specific mouse competition $m_m=0.5$. Each panel shows how each component of the equilibrium curves ($O^*$, $S^*$ and $M^*$, respectively) evolves with respect to $m$. Along the stable intervals, the equilibira are shown as blue solid curves; they are shown as a dashed curve if they have one unstable direction, and as a dotted curve for two unstable directions. The transcritical bifurcation as marked with colored diamonds. Different colors are used for the unstable portions of the equilibirum curve, in order to make each individual curve more easily identifiable across panels). For this simulation $m_m=0.5$, $o=1$, $s=1$ and the other fixed parameters are specified in Table~\ref{params}.}}
\label{bif_m_competition_mm_0_5}
\end{center}
\end{figure}

We therefore first investigated the effects of increasing mouse extermination rates (i.e., of decreasing the parameter $m$) under two scenarios, consisting of different levels of murine intra-specific competition $m_m$:  high ($m_m=0.5$) and low competition ($m_m=0.05$). Figure~\ref{bif_m_competition_mm_0_5} illustrates the results of varying $m$ for high competition $m_m$. The system's equilibria (computed and extended with the Matcont software) are shown in each figure panel from the perspective of a different variable/species ($O$, $S$ and $M$, respectively). As $m$ changes, these curves cross and swap stability via an interesting sequence of transcritical bifurcations. We notice that at each $m$ value there is only one locally attracting equilibrium branch (shown as a blue curve) which undergoes all these transitions. Following this curve we can document that, as $m$ decreases, the steady state population also decreases in all of its components -- with $O^*$, $S^*$, $M^*$ captured separately in panels (a)-(c). This appears as a naturally graded result, with perturbations in $m$ inducing linear changes in the steady state, with slope changes when switching between branches around bifurcation points. As expected, higher reproduction rates (and/or lower extermination rates) lead to higher asymptotic values in all three compartments of the system, as one would expect.

More specifically: as $m$ is increased from zero, the system has first a unique equilibirum corresponding to owl and snake extinction ($O^*=S^*=0$); the mouse asymptotic population increases linearly with $m$ until the trascritical bifurcation at $m^*=1.66$ (marked with an orange diamond). At the bifurcation point, the attractive equilibirum branch changes to one in which owl still go extinct, but both snakes and mice survive and increase with $m$, until the next bifurcation point $m^*=2.71$ (pink diamond). For $m>2.71$, all species survive, with asymptotic levels that increase linearly with $m$.

\begin{figure}[h!]
\begin{center}
\includegraphics[width=0.8\textwidth]{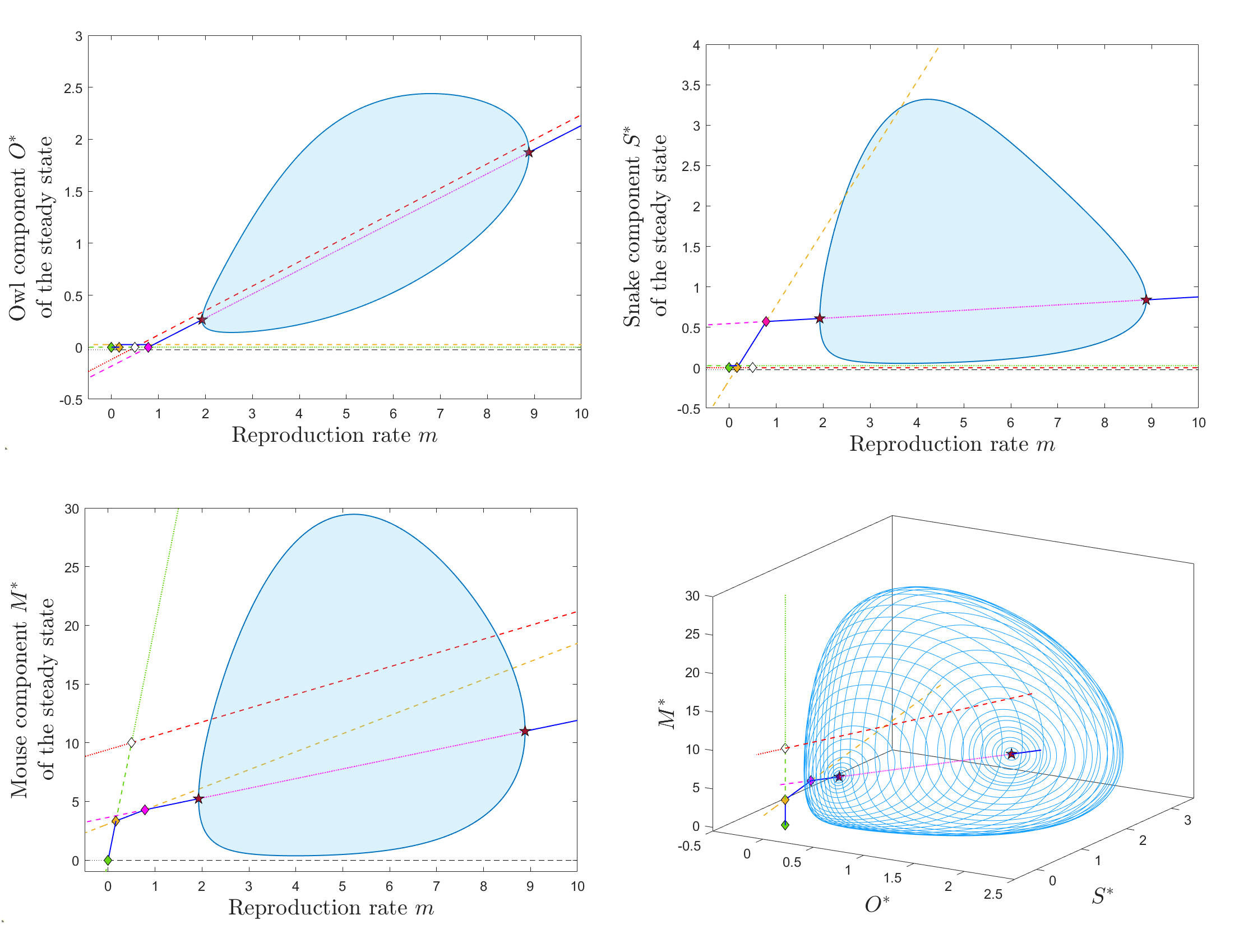}
\caption{\small \emph{{\bf Bifurcation diagram with respect to the mouse reproduction rate $m$}, for low intra-specific mouse competition $m_m=0.05$. Each panel shows how each component of the equilibrium curves ($O^*$, $S^*$ and $M^*$, respectively) evolves with respect to $m$. The equilibira are shown as blue solid curves along the stable intervals, as dashed curves if they have one unstable direction, and as dotted curves of they have two unstable directions. Transcritical bifurcations as marked with colored diamonds and subcritical Hopf bifurcations are shown as brown stars. For this simulation $m_m=0.05$, $o=1$, $s=1$ and the other fixed parameters are specified in Table~\ref{params}.}}
\label{bif_m_competition_mm_0_05}
\end{center}
\end{figure}

This scenario changes significantly, however, in the case of low intra-specific competition $m_m=0.05$. This is due primarily to the presence of stable cycles, for a parameter range delimited by two subcritical Hopf bifurcations, , as shown in Figure~\ref{bif_m_competition_mm_0_05}. This window of oscillations changes the scenario previously discussed in the high competition case, in which -- once the system has a non-extinction steady state -- increments in $m$ simply lead to higher values in all components of this steady state. In this case, increasing $m$
past the lower Hopf bifurcation point transitions the system into globally stable oscillations, with amplitude and geometry depending on the value of $m$. As $m$ is increased from zero, the system has first a unique equilibirum corresponding to owl and snake extinction ($O^*=S^*=0$); the mouse asymptotic population increases linearly with $m$ until the trascritical bifurcation at $m^*=0.15$ (marked with an orange diamond). At the bifurcation point, the attractive equilibirum branch changes to one in which owl still go extinct, but both snakes and mice survive and increase with $m$, until the next bifurcation point $m^*=0.78$ (pink diamond). If $m$ is increased passed the pink threshold value, the equilibrium undergoes a subcritical Hopf bifurcation at $m^*=1.92$, with creation of stable cycles. These cycles grow and then diminish in amplitude as $m$ continues to increase, and they disappear through a second Hopf bifurcation at $m^*=8.88$, where the stability of the equilibirum is regained. After the second Hopf bifurcation, the asymptotic levels continue to increase as $m$ increases.

The figure shows how the lowest values of the stable cycle is extremely close to zero in the $S$ and $M$ components for a relatively large interval around $m=4$. While the oscillations cannot terminate either species in and of themselves, they can render these two species very vulnerable to small perturbations. For example, for $m \sim 4$, the accidental death of a few mice or a few snakes may lead to extinction of the corresponding species, and collapse of the whole model. Interestingly, as the apex predator of the system, owls seem less exposed to small fluctuations (the cycle projections in panel (a) present with a bigger separation from zero). The cycling ends when $m$ transcends the higher Hopf value, after which the nonextinction stable equilibrium resumes its linearly increasing course.

This phenomenon is interesting, since it suggests that higher reproduction rates in the common pray don't not always promote a robust eco-system. In our example, rates $m$ in between the two Hopf points may endanger the integrity of the system, and should be avoided. There is a subsequent lesson in terms of human control and environmental preservation efforts: supporting the murine population when its rates are low, or exterminating it when the rates are high must be done with careful previous documentation and knowledge of the system's behavior, if one does not want to increase the eco-system's vulnerability. This is especially because of the additional dependence of such behaviors of the other system parameters (e.g. the inner competition $m_m$, in this case). 

\subsection{Dependence on owl reproduction rate $o$}

A similar phenomenon can be observed when varying the owl death rate $o$. (Recall that, since they are surviving on predation as only source of food, owls would die out in absence of the other two species, hence the negative term $-oO$.) In Figure~\ref{bif_o_competition_mm_0_05}, we show the evolution of the system's equilibria when the death rate $o$ is increased. As before, the three compartments of the equilibira and cycles are plotted separately in panels (a)-(c), but also together in panel (d).

\begin{figure}[h!]
\begin{center}
\includegraphics[width=0.8\textwidth]{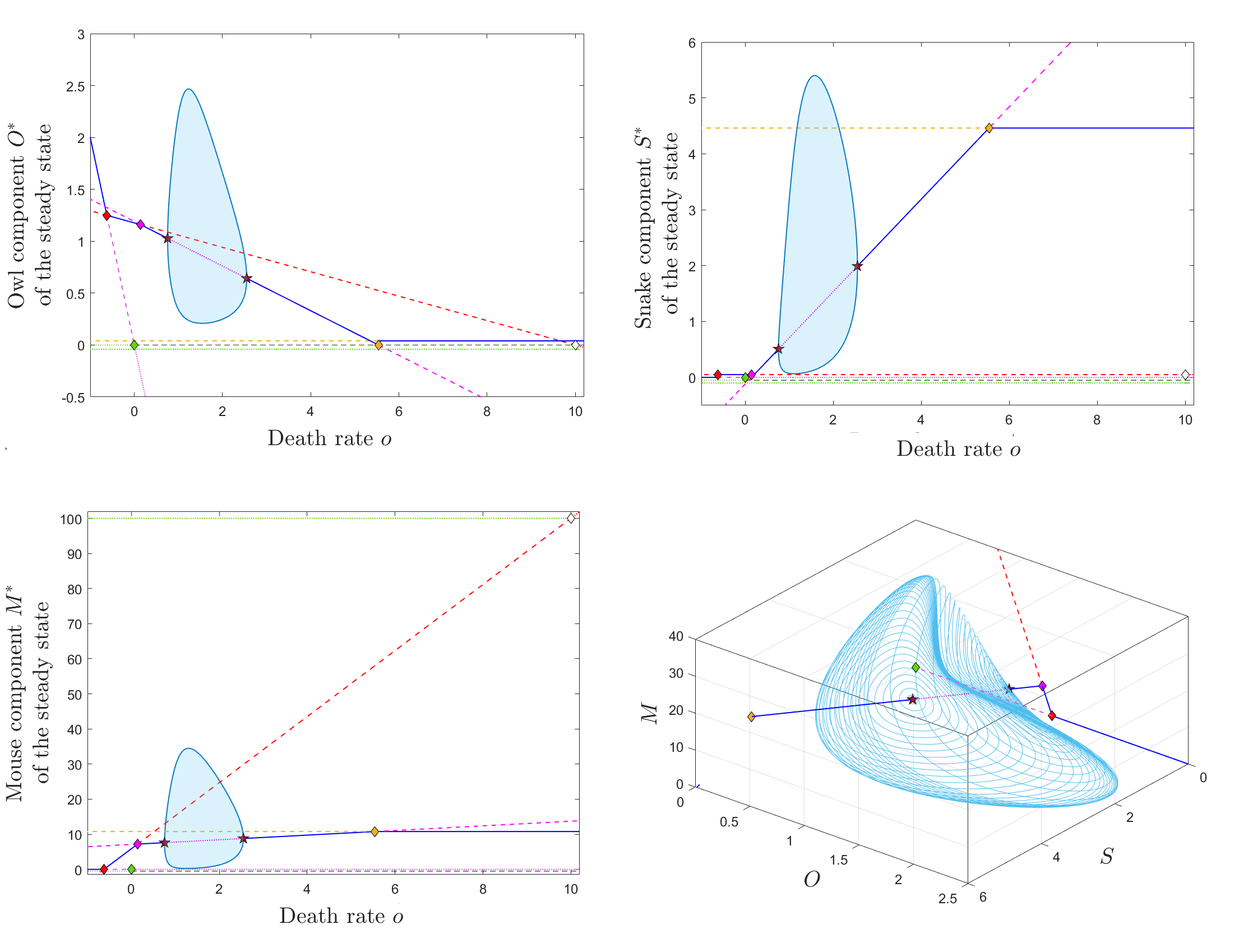}
\caption{\small \emph{{\bf Bifurcation diagram with respect to the owl reproduction rate $o$}, for the same intra-specific mouse competition $m_m=0.05$ as in Figure~\ref{bif_m_competition_mm_0_05}. Each panel shows how each component of the equilibrium curves ($O^*$, $S^*$ and $M^*$, respectively) evolves with respect to $o$. The equilibira are shown as blue solid curves along the stable intervals, as dashed curves if they have one unstable direction, and as dotted curves of they have two unstable directions. Transcritical bifurcations as marked with colored diamonds and subcritical Hopf bifurcations are shown as brown stars. For this simulation $m_m=0.05$, $m=5$, $s=1$ and the other fixed parameters are specified in Table~\ref{params}.}}
\label{bif_o_competition_mm_0_05}
\end{center}
\end{figure}

Specifically, the figure show that, as the death rate $o$ is increased from zero, the system has first a unique equilibirum corresponding to thriving owl population, but snake and mouse extinction ($M^*=S^*=0$). This situation only persists until $o^*=0.14$ (the transcrtitical bifurcation makes with a a pink diamond). The blue curve continues after the bifurcation with a non-extinction stable equilibrium, for which $O^*$ increases linearly with $o$, $S^*$ decrases linearly with $o$ and $M^*$ remains relatively stable. At $o^*=0.75$ (lower brown star), the system is thrown into stable oscillations by crossing a subcritical Hopf bifurcations. The oscillation are ended by a second Hopf bifurcation (higher brown star) at $o^*=2.54$. Beyond the higher Hopf bifurcation, the stable equilibirum is regained, with $O^*$ decaying to zero as $o$ decreases. After $o^*=5.53$ (orange transcritical bifurcation), the stability is picked up by an owl-extinction equilibirum branch (with the other two components at their corresponding carrying capacities $S^*=4.46$ and $M^*=10.76$).

In some ways, this evolution is not too surprising. If owl are hunted in excess, they will die out, leaving the snakes to take over as the apex (unique, in fact) predator. If the killing rate is lowered past the yellow bifurcation threshold, owl begin to recover, at the slight detriment of the snakes and even mice. This effect is linear, until the Hopf bifurcation point it hit, where lowering $o$ more triggers oscillations in the system. While there is nothing intrinsically unexpected or detrimental in cyclic behavior, here as well the lower values of the cycle are concerning, especially in the snake compartment. The values of $S$ become periodically so close to zero, that mere accidents may lead to species extinction. This remains the case until $o$ crosses the lower Hopf point, and stability of the equilibirum is restored. Even though the re-established equilibrium has now lower $S^*$ and $M^*$ values than before the Hopf window, the acute (if temporary) vulnerability rendered by oscillations is avoided. Of course, further lowering of $o$ will eventually lead to snake extinction; however, in this simulation, that only occurs as very low values of $o$ (between the red and the green diamonds. This path again suggests that one should not assume a monotone response from the system, and that in this case as well, previous knowledge of the cycling range would inform on how to control the values of $o$ in order to avoid the small region of exacerbated vulnerability.

\subsection{Dependence on snake reproduction rate $s$}

\begin{figure}[h!]
\begin{center}
\includegraphics[width=0.8\textwidth]{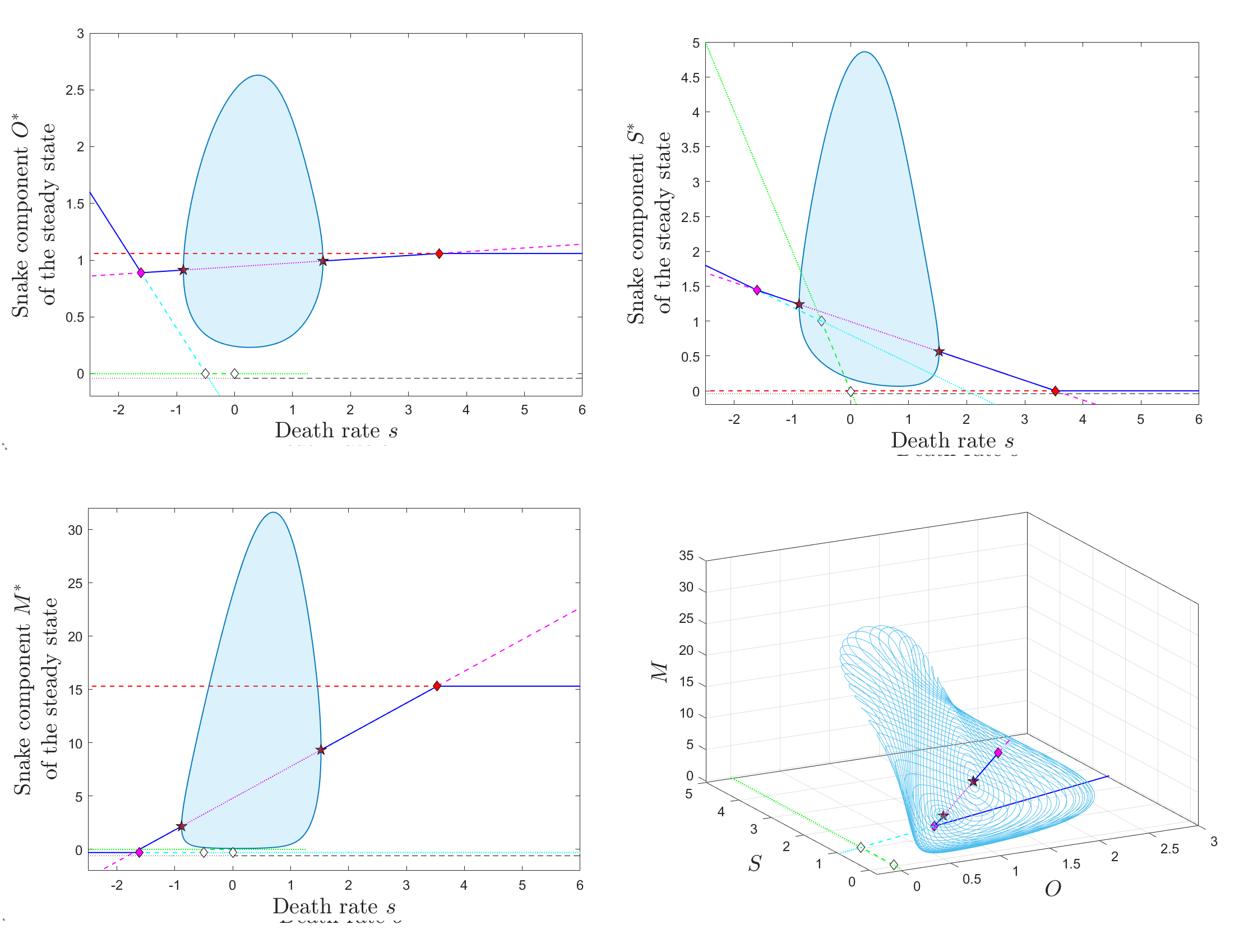}
\caption{\small \emph{{\bf Bifurcation diagram with respect to the owl reproduction rate $s$}, for intra-specific mouse competition $m_m=0.05$ as in Figure~\ref{bif_m_competition_mm_0_05}. Each panel shows how each component of the equilibrium curves ($O^*$, $S^*$ and $M^*$, respectively) evolves with respect to $s$. The equilibira are shown as blue solid curves along the stable intervals, as dashed curves if they have one unstable direction, and as dotted curves of they have two unstable directions. Transcritical bifurcations as marked with colored diamonds and subcritical Hopf bifurcations are shown as brown stars. For this simulation $m_m=0.05$, $m=5$, $o=1$ and the other fixed parameters are specified in Table~\ref{params}.}}
\label{bif_o_competition_mm_0_05}
\end{center}
\end{figure}

For completion, we will finally look at the effect of eliminating snakes from the system. Clearly, as expected, high killing rates $s$ will lead to extinction of snakes, and survival of the other two species, as a two-population predator-pray system. Notice that, for $s$ above the red transcritical point, $S^*=0$ and both $O^*$ and $M^*$ are constant under fluctuation in $s$. Once the rate is lowered beyond the red bifurcation, the snake population begins to recover, at the expense of the owls and mice (the steady states of which are starting to decrease linearly. This remains the case until the Hopf bifurcation is hit, and the system is prompted into oscillations, which periodically endanger both snakes and mice. In fact, for the parameter set at hand, the system will never exit the cycling regime as $s$ is lowered further, since the exit Hopf bifurcation occurs for a negative value of $s$. Therefore, a low positive $s$ will always create the vulnerability that comes with a cycle that passes too close to zero in two or the components ($S^*$ and $M^*$). Paradoxically, all populations are more likely to thrive for values of $s$ slightly higher than the cycling region.  

\subsection{Dependence on multiple parameters}

Of course, while the simulations in the previous sections illustrate possible scenarios as one parameter is varied at the time, one of the main points of our paper is to convey the idea that these scenarios may change significantly with the context (as encompassed by the other system parameters).

\begin{figure}[h!]
\begin{center}
\includegraphics[width=0.7\textwidth]{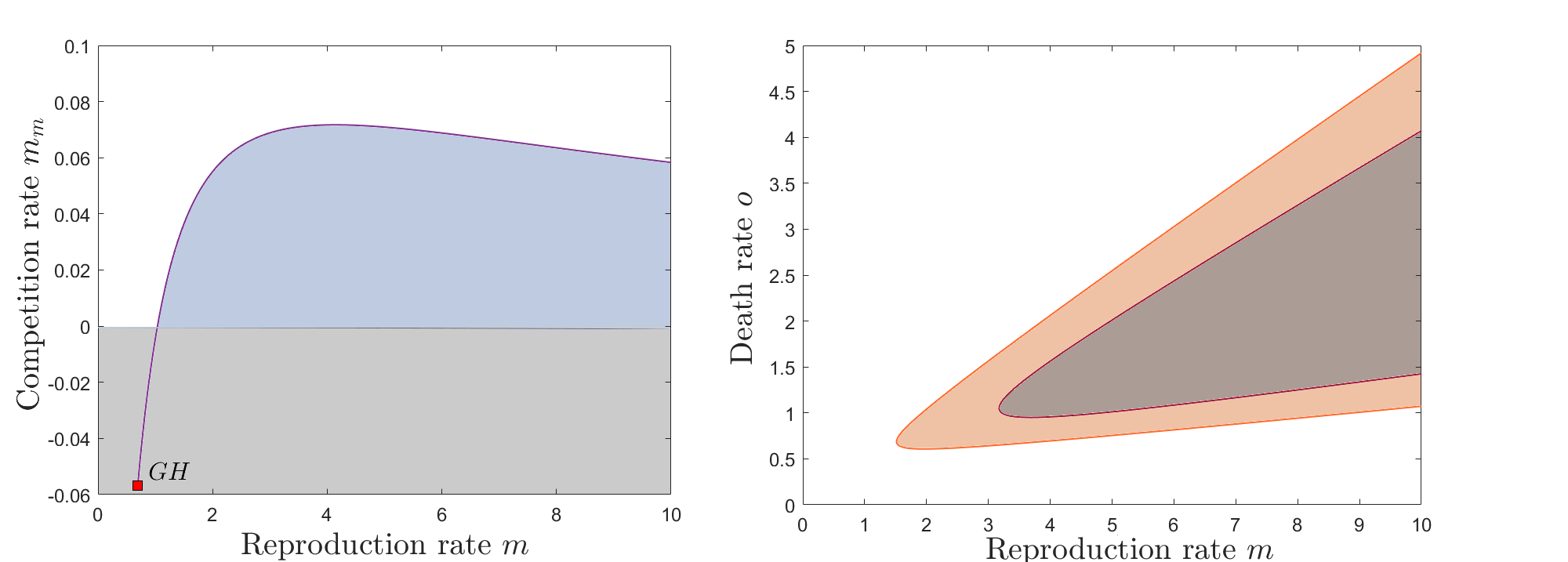}
\caption{\small \emph{{\bf Co-dimension two bifurcation diagrams}, showing the dependence of the system behavior on two different parameters at a time. {\bf Left.} Hopf curve in the $(m,m_m)$ parameter plane (purple curve) delimiting the region on stable cycling behavior (shaded in blue). The subcritical Hopf curve end in a codimension-two generalized Hopf point (red square) that lies in the unbiological half-plane $m_m<0$ (shaded in grey). For this figure, $o=s=1$, and the other parameters were fixed to their table values. {\bf Right.} Hopf curves in the $(m,o)$ parameter plane, for two different values of competition: $m_m=0.05$ (left, orange curve) and $m_m=0.07$ (right, brown curve). For this panel, $s=1$, and the other parameters were fixed to the table values.}}
\label{codim2}
\end{center}
\end{figure}

For example, Figure~\ref{codim2}a contextualizing the simultaneous dependence we observed both analytically and numerically on mouse reproduction and competition rates. Figure~\ref{bif_m_competition_mm_0_05}. When navigating the $(m,m_m)$ parameter plane, crossing the purple Hopf curve into the blue shaded region is equivalent to oscillation onset, and leaving the region stops these oscillations. Going beyond the two values of $m_m$ simulated earlier, the panel clarifies that: if $m<1.5$ (horizontal intercept of the Hopf curve), the system can't oscillate, irrespective of the value of $m_m$. When $m>1.5$, the system is oscillating for small values of $m_m$, and exits this regime as the $m_m$ increases and the Hopf curve is crossed. This is the distinction we observed in our simulations for $m=5$, when illustrating the asymptotic behavior for $m_m=0.05$ (underneath the Hopf curve) and $m-0.5$ (high above the Hopf curve, which is crossed at about $m_m=0.07$).

Figure~\ref{codim2}b throws a glimpse into the simultaneous dependence of the system on $m$ and $o$ (i.e., simultaneous control of mice and own rates). The parameter plane is divided by a Hopf curve into a colored region inside of the curve in which oscillations occur, and an outside regions where the system converges to a steady state. In particular, we can notice that, if $m$ is large enough to place us to the right of the curve turning point, increasing $o$ will always cross the shaded region, leading to a transition between two quantitatively different equilibria via a window of oscillations. In contrast, when $o$ is fixed above a certain threshold, increasing $m$ will cross into the oscillatory window, and never re-emerge, even with very high values of $m$. If one interprets oscillations as a situation that introduces a systemic vulnerability that needs to be avoided, one should not allow mice to reproduce wildly during periods of excessive owl hunting. We chose to illustrate the Hopf curve for two different values of $m_m$, to recall the point that the quantitative and even qualitative details of these scenarios may depend significantly on context.

\section{Discussion}

In this paper, we aimed to illustrate the importance of context when supporting or controlling and eco-system of predator-prey populations. As a working model, we used a less studied stratification with three interacting populations (owls, snakes and mice), in which the two predators (owls and snakes) have a common prey (mice), but such that the apex predator (owls) also preys on the intermediate predator (snakes). While we used these specific species as an example, we argued that the conceptualization can be used for any other model with similar relationships. 

For simplicity, we modeled all interactions based on Lotka-Volterra type equations, with linear and quadratic terms. We investigated this system both analytically and numerically, focusing primarily on understanding the effects on the overall system of controlling (exterminating) any of the three populations.

Our general expectation is that an increase in the reproduction rate of an eco-system's food supply (mice in our case) will produce proportional increases in the snake and owl populations. Similarly, it appears intuitive that trimming down the numbers of apex predators (owls) will lead to proportional increases in the snake and mice populations. According to this intuition, tightening pest control (extermination of mice) incrementally would only affect the system incrementally as well. In turn, small fluctuations in owl hunting practices cannot dramatically throw off the system's long-term prognosis.   
While this basic intuition is to some extent correct, there are significant cases where it fails.

As expected for a Lotka-Volterra system, we found parameter regions where the system converges to a stable cycle. This cycle's geometry is such that is visits periodically values that are dangerously close to zero in all compartments, snakes in particular. We interpreted this as a point of vulnerability for the system. While, if the cycle were to evolve deterministically, the population levels are set to recover periodically after each drop, random perturbations may have in this case a significant impact. Accidental death (by causes not included in this model) may endanger the species, when the size of the species is slimmed down to a few individuals (less than 10 snakes, in some cases). For parameter subsets where the cycle creates such bottle-necks, the system is exposed to total snake extinction, subsequent loss of a species and reduction to a two-species eco-system, which is a unwanted outcome. 

In this light, it becomes important to avoid the parameter subsets where such stable cycles occur. This goes against the intuition that all system components change monotonically with the reproduction/death parameters, and suggest that it may be desirable to decrease or even increase the extermination rate until the system exits the cycling behavior. More practically, one should not assume that fine tuning of a control parameter such as mouse extermination rate can only produce small perturbations in overall behavior. If operating near a bifurcation point, small changes may lead to significant long-term effects, and subject the system to increased vulnerability to external factors. To avoid unwanted outcomes, one need to be aware of these effects and trajectories ahead of time. This falls into a more general idea warning against making simplifying assumptions about the behavior of complex systems when planning on making practical decisions or implementing policies. It is precisely where mathematical modeling can be useful.

Finally, we would like to point out again that we built and analyzed a simplified model, with quadratic interactions. While this is convenient if one wants to preserve generality and tractability, studying more specific eco-systems would of course lead to finer mathematical assumptions, based on the particular interactions of the species at hand. Existing modeling research suggests that more complex nonlinear terms can lead to more complex transitions, and route to chaos~\cite{sen2018complex}.

\bibliographystyle{plain}
\bibliography{references}

\clearpage
\section*{Appendix: stability and Hopf bifurcations for the non-extinction equilibrium}

\noindent Necessary condition for Hopf bifurcation is in general:
\begin{equation}
P(\lambda) = (\lambda+\alpha)(\lambda^2+\beta^2) = \lambda^3+\alpha \lambda^2 + \beta^2 \lambda + \alpha \beta^2
\label{Hopf_cond}
\end{equation}

\noindent In our case, this translates to the following set of conditions on the parameters:
\begin{eqnarray}
&& o_oO^* + s_sS^* + m_mM^* = \alpha  \nonumber \\ 
&&(o_os_s+o_ss_o)O^*S^*+(o_om_m+o_mm_o)O^*M^*+(s_sm_m+s_mm_s)S^*M^* = \beta^2 \nonumber\\ 
&& O^*M^*S^*(o_os_sm_m-o_mm_ss_o+o_ss_ms_o+o_mm_os_s+m_ss_mo_o+s_oo_sm_m) = \alpha \beta^2
\label{Hopf_conds1}
\end{eqnarray}

\noindent To simplify the analysis, we introduce the following simplifying assumptions:

\begin{description}
\item[Assumption \#1.] The predation rates are equal, that is: owls are not biased towards eating mice versus snakes, and mice are equally preferred by snakes and by owls. This implies $o_s=o_m=s_m=p$;

\item[Assumption \#2.] The consumption incentive is the same for all predator-prey pairs, that is: mice are equally efficient in feeling owls and snakes, and snakes are an equally efficient food for owls as mice are. This implies $m_o=m_s=s_o=f$.

\item[Assumption \#3.] Competition is equally destructive within each species. This implies $o_o=s_s=m_m=c$.
\end{description}

\noindent Under these assumptions, the conditions for having an equilibrium at $(O^*,S^*,M^*)$ become:
\begin{eqnarray}
&& p(S^*+M^*) - cO^* = o \nonumber \\ 
&& pM^*-fO^* - cS^* = s \nonumber \\ 
&& cM^* + f(O^*+S^*) =m
\end{eqnarray}

\noindent and the Hopf bifurcation conditions~\eqref{Hopf_conds1} become:
\begin{eqnarray}
&& c(O^* + S^* + M^*) = \alpha  \nonumber \\ 
&&(pf+c^2)(O^*S^*+O^*M^*+S^*M^*) = \beta^2 \nonumber\\ 
&& (c^3+p^2f-pf^2+3cpf)O^*M^*S^* = \alpha \beta^2
\label{Hopf_conds2}
\end{eqnarray}

\noindent To further simplify computations, we introduce for now a fourth assumption, which we will later relax for a more general analysis;

\begin{description}
\item[Assumption \#4.] Predation and consumption rates are equal: $p=f$.
\end{description}

\noindent With this additional assumption, one can more easily compute the non-extinction equilibrium:
\begin{eqnarray}
O^* &=& \frac{-co(f^2+c^2) + fmc(c+f)-fsc(c-f)}{c^2+3f^2}\\ \nonumber \\
M^* &=& \frac{cm(f^2+c^2) + fsc(c+f)-foc(c-f)}{c^2+3f^2}\\ \nonumber \\
S^* &=& \frac{-cs(f^2+c^2) + foc(c+f)-fmc(c-f)}{c^2+3f^2}\\ \nonumber\\
\end{eqnarray}

\noindent We call $\delta = \frac{f}{c}$ the predation versus competition ratio, $\xi = m+s-o$, and introduce a change of variables as follows:
\begin{eqnarray*}
X = \frac{c^2+3f^2}{c^3} O^* &=& -o(\delta^2+1) + m\delta(\delta+1) -s\delta(1-\delta)   \\
&=& \delta^2(m+s-o) + \delta(m-s) - o = \delta^2 \xi + \delta(m-s) - o\\ \nonumber \\
Y = \frac{c^2+3f^2}{c^3} S^* &=& -s(\delta^2+1) + o\delta(\delta+1) +m\delta(1-\delta)   \\
&=& \delta^2(-m-s+o) + \delta(m+o) - s = -\delta^2 \xi + \delta(m+o) - s\\ \nonumber \\
Z = \frac{c^2+3f^2}{c^3} M^* &=& m(\delta^2+1) + s\delta(\delta+1) +o\delta(1-\delta)   \\
&=& \delta^2(m+s-o) + \delta(s+o) + m = \delta^2 \xi + \delta(s+o) + m\\ \nonumber \\
\end{eqnarray*}

The characteristic polynomial can be rewritten as:
\begin{eqnarray*}
P(\lambda) &=& \lambda^3 + c(O^*+M^*+S^*)\lambda^2+(c^2+f^2)(O^*M^*+O^*M^*+M^*S^*)\lambda + c(c^2+3f^2)O^*M^*S^* \\
&=& \lambda^3 + A\lambda^2+B\lambda+D
\end{eqnarray*}

\noindent so that:
\begin{eqnarray*}
A &=& c(O^*+M^*+S^*) = \frac{c^4}{c^2+3f^2} (X+Y+Z) = \frac{c^2}{1+3\delta^2} \Sigma_1\\\\
B &=& (c^2+f^2)(O^*M^*+O^*M^*+M^*S^*) = \frac{c^6 (c^2+f^2)}{(c^2+3f^2)^2} (XY+XZ+YZ) = \frac{c^4(\delta^2+1)}{(1+3\delta^2)^2} \Sigma_2 \\\\
D &=& c(c^2+3f^2)O^*M^*S^* = \frac{c^{10}(c^2+3f^2)}{(c^2+3f^2)^3} XYZ = \frac{c^6}{(1+3\delta^2)^2} \Sigma_3
\end{eqnarray*}

\noindent where the sigmas denote respectively $\Sigma_1 = X+Y+Z$, $\Sigma_2 = XY+XZ+YZ$ and $\Sigma_3 = XYZ$. Assuming positivity of the equilibrium, we have $\Sigma_i >0$, and subsequently $A,B,D>0$. Since the eigenvalues $\lambda_i$ satisfy the conditions $\lambda_1 + \lambda_2 + \lambda_3 = -A<0$, $\lambda_1\lambda_2 + \lambda_1 \lambda_3 + \lambda_2 \lambda_3 = B>0$ and $\lambda_1 \lambda_2 \lambda_3 = -C<0$, it follows that at least one eigenvalue is real and negative.\\

\noindent The Hopf condition~\eqref{Hopf_cond} requesting that $AD=B$ becomes: 
\begin{equation}
(1+\delta^2) \Sigma_1 \Sigma_2 = (1+ 3\delta^2) \Sigma_3
\end{equation}

\end{document}